\documentclass[monophys,vecphys]{svmono}
\usepackage{makeidx}         % allows index generation
\usepackage{graphicx}        % standard LaTeX graphics tool
\usepackage{multicol}        % used for the two-column index
\usepackage{bm}
\usepackage[bottom]{footmisc}% places footnotes at page bottom
\makeindex             % used for the subject index
\begin{document}
%frontmatter%%%%%%%%%%%%%%%%%%%%%%%%%%%%%%%%%%%%%%%%%%%%%%%%%%%%%%
%\include{dedic}
%\include{pref}
%\tableofcontents
%\mainmatter%%%%%%%%%%%%%%%%%%%%%%%%%%%%%%%%%%%%%%%%%%%%%%%%%%%%%%%
%\include{part}

%\include{ch_vulpiani} % Chapter Vulpiani et al.

\chapter{Role of chaos for the validity of
statistical mechanics laws: diffusion and conduction}

\chaptermark{The role of chaos for the validity of
statistical mechanics}

\vspace{-3.cm}
\section*{
\begin{center} 
   Massimo Cencini$^{1}$, Fabio Cecconi$^1$, \\ 
   Massimo Falcioni$^2$ and Angelo Vulpiani$^2$
\end{center}}

\subsubsection*{
\begin{center}
$^1$ INFM and Istituto dei Sistemi Complessi (ISC-CNR) \\
             Via dei Taurini 19, I-00185 Roma (Italy). \\
\vspace{0.3cm} $^2$ 
INFM and Dipartimento di Fisica Universit\`a ``La Sapienza'' \\ 
             P.le Aldo Moro 2, I-00185 Roma (Italy)  
\end{center}}

\begin{abstract}
Several years after the pioneering work by Fermi Pasta and Ulam,
fundamental questions about the link between dynamical and statistical
properties remain still open in modern statistical mechanics.
Particularly controversial is the role of deterministic chaos for the
validity and consistency of statistical approaches.  This contribution
reexamines such a debated issue taking inspiration from the problem of
diffusion and heat conduction in deterministic systems. Is microscopic
chaos a necessary ingredient to observe such macroscopic phenomena?
\end{abstract}

%%%%%%%%%%%%%%%%%%%%%%%%%%%%%%%%%%%%%%%%%%%%%%%%%%%%%%%%%%%%%%%%%%%%%%%%%%
\section{Introduction}
\label{sec:intro}
%%%%%%%%%%%%%%%%%%%%%%%%%%%%%%%%%%%%%%%%%%%%%%%%%%%%%%%%%%%%%%%%%%%%%%%%%%
Statistical mechanics, founded by Maxwell, Boltzmann and Gibbs, aims to
explain the macroscopic properties of systems with a huge number of
degrees of freedom without specific assumptions on the microscopic
dynamics, a part from ergodicity ~\cite{ehren,Galla}.  The discovery
of deterministic chaos \cite{ER}, beyond its undoubted important
implications on many natural phenomena, enforced us to reconsider some
basic problems standing at the foundations of statistical mechanics
such as, for instance, the applicability of a statistical description
to low dimensional systems.  However, even after many years, the
experts do not agree yet on the basic conditions which should ensure
the validity of statistical mechanics.
 
The spectrum of viewpoints found in literature is rather wide, ranging
from the Landau (and Khinchin~\cite{khin}) earlier belief on the key
role of the many degrees of freedom and the (almost) complete
irrelevance of ergodicity, to the opinion of  those who, 
as Prigogine and his
school~\cite{prig1} considers chaos as the crucial requirement to
develop consistent statistical approaches.  Recently some authors
(e.g. Lebowitz~\cite{lebo}, Bricmont~\cite{Bricmont}) have given new
life to the debate~\cite{driebe,prig1}, renewing the intuition of
Boltzmann~\cite{cerci} and Maxwell~\cite{garber} on the relevance of
the huge number of particles in macroscopic systems.

This volume offers the opportunity to celebrate the 100th and 50th
anniversaries of two of the most influential works in statistical
physics: Einstein's work on Brownian motion (1905) \cite{e1905} and
Fermi's one (1955) on the nonlinear chain of oscillators ({\it al
secolo} the FPU work, from the authors E.~Fermi, J.~Pasta and S.~Ulam
\cite{fpu}).  We shall discuss some aspects related to diffusion
problems and heat conduction focusing on the role of (microscopic)
chaos for the occurrence and robustness of these (macroscopic)
phenomena.  Transport phenomena, despite their ubiquity in everyday
life, are still subject of debate among theoretic physicists.

Because of the variety of specific interactions and technical
difficulties in realistic systems, simplified microscopic models are
unavoidable tools for the study of transport mechanism.  Several
simulations and theoretical works have shown that, in systems with very
strong chaos (namely hyperbolic systems), there exists a close
relationship between transport coefficients (e.g. viscosity,
diffusivity, thermal and electrical conductivity) and indicators of
chaos (Lyapunov exponents, KS entropy, escape rates) ~\cite{gas,dorf}.
At a first glance, the existence of such relations would support the
point of view of who considers chaos as the basic ingredient for the
applicability of statistical mechanics.  However, it is not possible
to extend those results to generic systems. In fact, we shall see that
many counterexamples prove that chaos is not a necessary condition for
the emergence of robust statistical behaviors~\cite{ford2,R3}.  In
particular, we shall see that phenomena such as diffusion \cite{DC}
and heat conduction \cite{LiWangHu} may take place also in non-chaotic
systems.  These and many other examples provide indication that
microscopic chaos is not the unique possible origin of macroscopic
transport in dynamical systems.

The material is organized into two, almost self-contained, parts. In
the first, after a brief historical introduction to the different
microscopic models proposed to explain macroscopic diffusion, we
discuss a recent experiment (and the consequent debate it stimulated)
aimed to prove that microscopic chaos is at the origin of Brownian
motion.  This gives us the possibility to introduce and discuss the
problem of diffusion in non-chaotic deterministic systems, and to
point point out the necessary microscopic conditions to observe
diffusion. The second part is mostly devoted to a discussion of the
celebrated FPU numerical experiments and its consequences for the
ergodic problem and heat conduction. We shall see that there are
non-chaotic models displaying (macroscopic) heat conduction,
confirming the non-essential role of chaos on transport.

%%%%%%%%%%%%%%%%%%%%%%%%%%%%%%%%%%%%%%%%%%%%%%%%%%%%%%%%%%%%%%%%%%%%%%%%%%
\section{On the microscopic origin of macroscopic diffusion}
\label{sec:diffusion} 
%%%%%%%%%%%%%%%%%%%%%%%%%%%%%%%%%%%%%%%%%%%%%%%%%%%%%%%%%%%%%%%%%%%%%%%%%%
At the beginning of the twentieth century, the atomistic theory of
matter was not yet fully accepted by the scientific community.  While
searching for phenomena that would prove, beyond any doubt, the
existence of atoms, Einstein realized that ``{\it ... according to the
molecular-kinetic theory of heat, bodies of microscopically-visible
size suspended in a liquid will perform movements of such magnitude
that they can be easily observed in a microscope ...}'', as he wrote
in his celebrated paper in 1905 ~\cite{e1905}.  In this work, devoted
to compare the different predictions that classical thermodynamics and
molecular-kinetic theory of heat make about those small bodies,
Einstein argued that their motion has a diffusive character. Moreover,
he discovered an important relation among the diffusion
coefficient $D$, the fluid viscosity $\eta$, the particles radius $a$
(having assumed spherical particles), Avogadro's number $N_A$, the
temperature $T$ and the gas constant $R$:
\begin{equation}
D = \frac{1}{N_A}  \frac{RT}{6\pi \eta a} \,\, .
\label{diff}
\end{equation}
Einstein relation (\ref{diff}), that may be seen as the first example
of the fluctuation-dissipation theorem~\cite{kubo}, allowed for the
determination of Avogadro's number~\cite{chandra} and gave one of the
ultimate evidences of the existence of atoms.
 
Einstein's work on Brownian motion (BM) is based on statistical
mechanics and thermodynamical considerations applied to suspended
particles, with the assumption of velocity decorrelation.
  
One of the first successful attempts to develop a purely dynamical
theory of BM dates back to Langevin~\cite{lange} that, as himself
wrote, gave ``{\it ... a demonstration} [of Einstein results] {\it
that is infinitely more simple by means of a method that is entirely
different.}''  Langevin considered the Newton equation for a small
spherical particle in a fluid, taking into account that the Stokes
viscous force it experiences is only a mean force. In one direction,
say e.g. the $x$-direction, one has:
\begin{equation}
m \,\, \frac{d^2 x}{dt^2} = -6 \pi \eta a \, \, 
\frac{d x}{dt} + F
\label{langevin}
\end{equation} 
where $m$ is the mass of the particle.  The first term of the
  r.h.s.  is the Stokes viscous force. The second one $F(t)$ is a
  fluctuating random force, independent of $v= dx/ dt$, modeling the
  effects of the huge number of impacts with the surrounding fluid
  molecules, which is taken as a zero-mean, Gaussian process with
  covariance $\langle F(t) F(t') \rangle = c \delta(t-t')$. The
  constant $c$ is determined by the equipartition condition $\langle
  (dx/dt)^2\rangle=RT/(mN_A)$, i.e.  $c= 12\pi\eta a RT/N_A$.
 
Langevin's work along with that of Ornstein and Uhlenbeck~\cite{uhle}
are at the foundation of the theory of stochastic differential
equations. The stochastic approach is however unsatisfactory
being a phenomenological description.
 
The next theoretical challenge toward the building of a dynamical
theory of Brownian motion is to understand its microscopic origin from
first principles. Almost contemporaneously to Einstein's efforts,
  Smoluchowski tried to derive the large scale diffusion of Brownian
  particles from the similar physical assumptions about their
  collisions with the fluid molecules~\cite{Smoluch}.
  
A renewed interest on the subject appeared some years later, when it
was realized that even purely deterministic systems composed of a
large number of particles give rise to macroscopic diffusion, at least
on finite time scales.  These models had an important impact in
justifying Brownian motion theory and, more in general, in deriving a
consistent microscopic theory of irreversibility.
 
Some of these works considered chains of harmonic oscillators of equal
masses~\cite{turner1,mazur,ford1,phill}, while
others~\cite{rubin,mazur2} analyzed the motion of a heavy impurity
linearly coupled to a chain of equal mass oscillators.  When the
number of oscillators goes to infinity, the momentum of the heavy
particle was proved to behave as a genuine stochastic process
described by the Langevin equation~(\ref{langevin}).  When their
number is finite, diffusion remains an effective phenomenon lasting
for a (long but) finite time.
 
Soon after the discovery of dynamical chaos \cite{Lorenz}, it was
realized that also simple low dimensional deterministic systems may
exhibit a diffusive behavior.  In this framework, the two-dimensional
Lorentz gas~\cite{lorentz}, describing the motion of a free particle
through a lattice of hard round obstacles, provided the most valuable
example.  As a consequence of the obstacle convexity, particle
trajectories are chaotic, i.e., aside from a set initial conditions of
zero measure, exhibit a positive and finite Lyapunov exponent,.  At
long times, for the case of billiards, the mean squared displacement
from the particle initial position grows linearly in time.  A Lorentz
system with periodically arranged scatterers is closely related to the
Sinai billiard~\cite{sinai1,sinai2}, which can be obtained from the
former by folding the trajectories into the unitary lattice cell. The
extensive study on billiards has shown that chaotic behavior might
usually be associated with diffusion in simple low dimensional models,
supporting the idea that chaos was at the very origin of diffusion.
However, more recently (see e.g. Ref.~\cite{DC}) it has been shown
that even non-chaotic deterministic systems, such as a bouncing
particle in a two-dimensional billiard with polygonal but randomly
distributed obstacles (wind-tree Ehrenfest model), may exhibit a
diffusion-like properties. This example can lead to think that the
external source of randomness may play a role similar to chaos (for a
more detailed discussion about this point see
Sect.~\ref{sec:nonchaotic}).
 
Deterministic diffusion is a generic phenomenon present
also in simple chaotic maps on the line.  Among the many contributions
we mention the work by Fujisaka, Grossmann~\cite{fuji,gross} and
Geisel~\cite{gei1,gei2}.  A typical example is the
one-dimensional discrete-time dynamical system:
\begin{equation}
x(t+1) = [x(t)] + F(x(t)-[x(t)]) \,  ,
\label{eq:chaos}
\end{equation}
where $x(t)$ (the position of a point-like particle) performs
diffusion in the real axis.  The bracket $[\dots]$ denotes the integer
part of the argument. $F(u)$ is a map defined on the interval $[0,1]$
that fulfills the following requirements:\\ \indent i) The map,
$u(t+1)=F(u(t))$ (mod $1$) is chaotic.\\ \indent ii) $F(u)$ must be
larger than $1$ and smaller than $0$ for some values of $u$, so to
have a non vanishing probability to escape from each unit cell (a unit
cell of real axis is every interval $C_{\ell} \equiv [\ell,\ell+1]$,
with $\ell \in {\bm Z}$).\\ \indent iii) $F_r(u)=1-F_l(1-u)$, where
$F_l$ and $F_r$ define the map in $u\in [0,1/2[$ and $u\in [1/2,1]$
respectively. This anti-symmetry condition with respect to $u= 1/2$ is
introduced to avoid a net drift.
 
A very simple and much studied example of $F$ is 
\begin{eqnarray}
F(u) =
\left\{
\begin{array}{ll}
2(1+a) u   \qquad \qquad \quad \, \, \mbox{if~}  u \in [0,1/2[   \\
2(1+a) (u-1) + 1 \,\quad\mbox{if~}  u \in [1/2,1]
\end{array}\right.
\label{eq:chaos2}
\end{eqnarray}
where $a>0$ is the control parameter.  It is useful to remind the link
between diffusion and velocity correlation, i.e. the Taylor-Kubo
formula, that helps to unravel how diffusion can be realized in
different ways.  The velocity correlation function is defined as
$C(\tau)=\langle v(\tau) v(0) \rangle$, where $v(t)$ is the velocity
of the particle at time $t$.  It is easy to see that for continuous
time systems (e.g. Eq.(\ref{langevin}))
\begin{equation}
\langle \left[ x(t)-x(0)\right]^2 \rangle 
\simeq  2\; t\; \int_0^t 
d\tau \; C(\tau)\,.
\label{pippo1}
\end{equation}
Standard diffusion, with  $D = \int_0^{\infty} \; d \tau \;
C(\tau)$, is always obtained whenever the hypotheses for the
validity of the central limit theorem are verified: \\
$i)$ finite variance of the velocity: $\langle v^2 \rangle < \infty$;\\
\noindent$ii)$ faster than $\tau^{-1}$ decay of the velocity
correlation function $C(\tau)$\footnote{In discrete-time systems, the
velocity $v(t)$ and the integral $\int d\tau C(\tau)$ are replaced by
the finite difference $x(t+1) - x(t)$ and by the quantity $\langle
v(0)^2 \rangle/2 + \sum_{\tau \geq 1} C(\tau)$, respectively.}.

The first condition, independently of the microscopic dynamics under
consideration (stochastic, deterministic chaotic or regular), excludes
unphysical models, i.e. with infinite variance for the velocity.  The
second requirement corresponds to a rapid memory loss of initial
conditions.  It is surely verified for the Langevin dynamics where the
presence of the stochastic force entails a rapid decay of
$C(\tau)$. In deterministic regular systems, such as the model of
  many oscillators, the velocity decorrelation (i.e., the small
fluctuations of $C(\tau)$ around zero, for almost all the time) is the
result of the huge number of degrees of freedom that act as a heat
bath on a single oscillator. In the (non-chaotic) Ehrenfest wind-tree
model decorrelation originates from the disorder in the obstacle
positions.  Deterministic chaotic systems, in spite of the fact that
nonlinear instabilities generically lead to a memory loss, are more
subtle. Indeed, there are many examples, namely intermittent
systems~\cite{R1}, characterized by a slow decay of the velocity
correlation function.

We end this section by asking whether it is possible to determine, by
the analysis of a Brownian particle, if the microscopic dynamics
underlying the observed macroscopic diffusion is stochastic,
deterministic chaotic or regular.

%%%%%%%%%%%%%%%%%%%%%%%%%%%%%%%%%%%%%%%%%%%%%%%%%%%%%%%%%%%%%%%%%
\subsection{Chaos or Noise? A difficult dilemma}
%%%%%%%%%%%%%%%%%%%%%%%%%%%%%%%%%%%%%%%%%%%%%%%%%%%%%%%%%%%%%%%%%
Inferring the microscopic deterministic character of Brownian motion
on an experimental basis would be attractive from a fundamental
viewpoint. Moreover it could provide further evidence to some recent
theoretical and numerical studies~\cite{PH88,BDPD97}.  Before
discussing a recent experiment\cite{gaspard} in this direction, we
must open the ``Pandora box'' of the longstanding and controversial
problem of distinguishing chaos from noise in signal
analysis~\cite{cenc} (see also
Refs.~\cite{Kantzbook,Abarnabel,Sugihara90a,Casdagli91,Kaplan92a,Kaplan93c}).
For the sake of clearness on the terminology used here, we specify
that: ``chaos'' refers to the motions originating from a deterministic
system with at least one positive but finite Lyapunov exponent and
therefore a positive and finite Kolmogorov-Sinai entropy; ``noise''
instead denotes the outcomes of a continuous valued stochastic process
with infinite value of Kolmogorov-Sinai entropy.

The first observation concerning the chaos/noise distinction is that,
very often in the analysis of experimental time series, there is not a
unique model of the ``system'' that produced the data. Moreover, even
the knowledge of the ``true'' model might not be an adequate answer
about the character of the signal. From this point of view, BM is a
paradigmatic example: in fact it can be modeled by a stochastic as
well as by a deterministic chaotic or regular process.

In principle a definite answer exists. If we were able to determine the
maximum Lyapunov exponent ($\lambda$) or the Kolmogorov-Sinai entropy
($h_{KS}$) of a data sequence, we would know without uncertainty
whether the sequence was generated by a deterministic law ($\lambda ,
h_{KS} < \infty$) or by a stochastic one ($h_{KS} \to
\infty$). Nevertheless, there are unavoidable practical limitations in
computing such quantities. They are indeed defined as infinite time
averages taken in the limit of arbitrary fine resolution. But, in
experiments, we have access only to a finite, and
often very limited, range of scales and times.
 
However, there are measurable quantities that are appropriate for
extracting meaningful information from the signal. In particular, we
shall consider the $(\epsilon,\tau)$-entropy per unit
time~\cite{kolmogorov,shannon,Gaspard93c} $h(\epsilon,\tau)$ that
generalizes the Kolmogorov-Sinai entropy (for details see next section
Cfr. Eq.~(\ref{eq:2-3b})).  In a nutshell, while for evaluating
$h_{KS}$ one has to detect the properties of a system with infinite
resolution, for $h(\epsilon, \tau)$ a finite scale (resolution)
$\epsilon$ is involved. The Kolmogorov-Sinai entropy is recovered in
the limit $\epsilon \to 0$, i.e. $h(\epsilon,\tau) \to h_{KS}$. This
means that if we had access to arbitrarily small scales, we could
answer the original question about the character of the law that
generated the recorded signal. Even if this limit is unattainable,
still the behavior of $h(\epsilon,\tau)$ provides a very useful
scale-dependent description of the signal character \cite{cenc,rass}.

For instance, chaotic systems ($0< h_{KS} < \infty$) are typically
characterized by $h(\epsilon,\tau)$ attaining a plateau $\approx
h_{KS}$, below a resolution threshold, $\epsilon_c$, associated with
the smallest characteristic length scale of the system.  Instead, for
$\epsilon>\epsilon_c$ $h(\epsilon,\tau)<h_{KS}$, and in this range the
details of the $\epsilon$-dependence may be informative on the large
scale (slow) dynamics of the system (see
e.g. Refs.~\cite{cenc,rass}). Indeed, at large scales typically
chaotic systems give rise to behaviors rather similar to stochastic
processes (e.g. the diffusive behavior discussed in the previous
subsection) with characteristic $\epsilon$-entropy. In stochastic
signals,  although $h_{KS}=\infty$, for any $\epsilon>0$,
$h(\epsilon,\tau)$ is a finite function of $\epsilon$ and $\tau$. The
dependence of $h(\epsilon,\tau)$ on $\epsilon$ and $\tau$, when known,
provides a characterization of the underlying stochastic process (see
Refs.~\cite{kolmogorov,Gaspard93c}).  For instance, stationary
Gaussian processes with a power spectrum\footnote{The power
spectrum $S(\omega)$ is the Fourier transform of $\langle
[x(t)-x(0)]^2\rangle$} $S(\omega)\propto \omega^{-(2\alpha+1)}$
(being $0 < \alpha <1$) are characterized by a power-law
$\epsilon$-entropy~\cite{kolmogorov}:
\begin{equation}
\lim_{\tau\to 0} h(\epsilon,\tau) \sim {\epsilon^{-1/\alpha}} \; .
\label{eq:kolmo56}
\end{equation}
The case $\alpha=1/2$, corresponding to the power spectrum of a
Brownian signal, would give $h(\epsilon)\sim \epsilon^{-2}$. Other
stochastic processes, such as {\em e.g.} time uncorrelated and bounded
ones, are characterized by a logarithmic divergence below a critical
scale, $\epsilon_c$, which may depend on $\tau$.

%%%%%%%%%%%%%%%%%%%%%%%%%%%%%%%%%%%%%%%%%%%%%%%%%%%%%%%%%%%%%%%%%%%%%%
\subsubsection{Definition and computation of the $\epsilon$-entropy}
%%%%%%%%%%%%%%%%%%%%%%%%%%%%%%%%%%%%%%%%%%%%%%%%%%%%%%%%%%%%%%%%%%%%%%
For the sake of self-consistency in this subsection we provide some basic
information on the definition and measurement (from experimental data) of the
$\epsilon$-entropy, which was originally introduced in the context of
information theory by Shannon~\cite{shannon} and, later, by
Kolmogorov~\cite{kolmogorov} in the theory of stochastic processes.
The interested reader may find more details in \cite{Gaspard93c} 
and \cite{berger}.

An operative definition of $h(\epsilon,\tau)$ is as follows.
Given the time evolution of a continuous variable ${\bf x}(t) \in \Re^d$, that
represents the state of a $d$-dimensional system, one introduces
the vector in $\Re^{md}$
\begin{equation}
\label{eq:2-1}
{\bf X}^{(m)}(t)= \left( {\bf x}(t), \dots, 
{\bf x}(t+m\tau-\tau) \right)\,, 
\end{equation}
which represents a portion of the trajectory, sampled at a discrete
time interval $\tau$.  After partitioning the phase space $\Re^d$
using hyper-cubic cells of side $\epsilon$, ${\bf X}^{(m)}(t)$ is
coded into a $m$-word: $W^{m}(\epsilon,t)=[i(\epsilon, t), \dots,
  i(\epsilon, t+m \tau -\tau)]$, where $i(\epsilon, t+j \tau)$ labels
the cell in $\Re^d$ containing ${\bf x}(t+j \tau)$. For bounded
motions, the number of available cells (i.e.  the alphabet) is
finite. Under the hypothesis of stationarity, the probabilities
$P(W^{m}(\epsilon))$ of the admissible words $\lbrace W^{m}(\epsilon)
\rbrace$ are obtained from the time evolution of ${\bf X}^{(m)}(t)$.
Then one introduces the $m$-block entropy, $H_m (\epsilon,\tau) = -
\sum _{ \lbrace W^{m}(\epsilon) \rbrace } P(W^{m}(\epsilon)) \ln
P(W^{m}(\epsilon))$, and the quantity $h_m(\epsilon , \tau)= \lbrack
H_{m+1} (\epsilon,\tau) -H_m (\epsilon,\tau) \rbrack/\tau$.  The
$(\epsilon,\tau)$-entropy per unit time, $h(\epsilon , \tau)$ is
defined by~\cite{shannon}:
\begin{equation}
\label{eq:2-3b}
h(\epsilon , \tau)=\lim _{m \to \infty} h_m(\epsilon , \tau) \,.
\end{equation} 
The Kolmogorov-Sinai entropy is obtained in the limit of small
$\epsilon$
\begin{equation}
\label{eq:2-5}
h_{KS} = \lim_{\epsilon \to 0} h(\epsilon, \tau)\;.
\end{equation} 
In principle, in deterministic systems $h(\epsilon)$, and henceforth
$h_{KS}$, does depend neither on the sampling time $\tau$~\cite{ER},
nor on the chosen partition because its rigorous
definition\cite{Gaspard93c} would require the infimum to be taken over
all possible partitions with elements of size smaller than
$\epsilon$. However, in practical computations, the specific value of
$\tau$ is important, and the impossibility to take the infimum over
all the partitions implies that, at finite $\epsilon$, $h(\epsilon)$
may depend on the chosen partition.  Nevertheless, for small
$\epsilon$, the correct value of the Kolmogorov-Sinai entropy is
usually recovered independently of the partition \cite{ER}.

Let us stress that partitioning the phase space does not mean a
discretization of the states of the dynamical system, which still
evolves on a continuum. The partitioning procedure corresponds to a
coarse-grained description (due, for instance, to measurements
performed with a finite resolution), that does not change the
dynamics. On the contrary, discretizing the states would change the
dynamics, implying periodic motions in any deterministic systems.
This happens, for instance, in any floating point computer
simulations; however such periods are, apart from trivial cases, very
long and practically undetectable.

In experimental signals, usually, only a scalar variable $u(t)$
can be measured and moreover the dimensionality of the phase space is
not known a priori.  In these cases one uses delay embedding
techniques~\cite{Kantzbook,Abarnabel}, where the vector ${\bf
  X}^{(m)}(t)$ is build as $\left( u(t), u(t+\tau), \dots,
u(t+m\tau-\tau) \right)$, now in $\Re^m$. This is a special instance
of (\ref{eq:2-1}). Then to determine the entropies $H_m(\epsilon)$,
very efficient numerical methods are available~(the reader may find an
exhaustive review in Ref.~\cite{Kantzbook}). The delay embedding
  procedure can be applied to compute the $\epsilon$-entropy of 
  deterministic and stochastic signals as well. The dependence of the
  $\epsilon$-entropy on the observation scale $\epsilon$ can be used
  to characterize the process underlying the
  signal~\cite{Gaspard93c}.

%------------------------Fig. 1 ---------------------------------------
\begin{figure}
\centering
\includegraphics[draft=false, scale=0.8, clip=true]{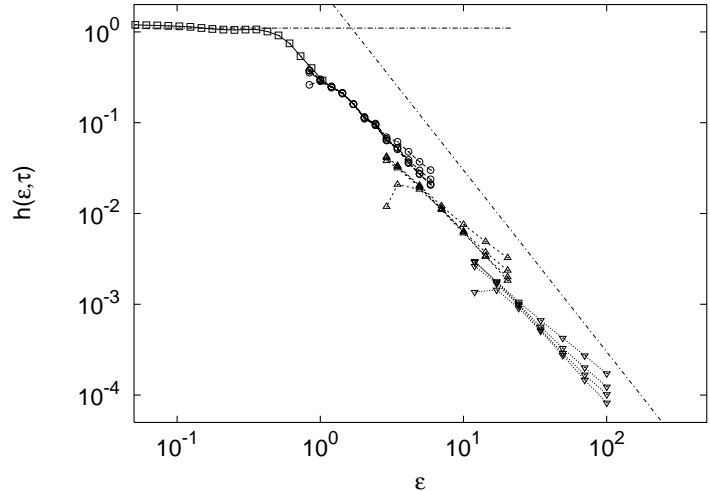}
\caption{ Numerically evaluated $(\epsilon,\tau)$-entropy for the map
(\ref{eq:mappa}) with $p=0.8$ computed by the standard techniques
\cite{Kantzbook} at $\tau=1$ ($\circ$), $\tau=10$ ($\bigtriangleup$)
and $\tau=100$ ($\bigtriangledown$) and different block length
($m=4,8,12,20$).  The boxes refer to the entropy computed with
$\tau=1$ but by using periodic boundary condition over $40$
cells. The use of periodic boundary conditions is necessary
to probe scales small enough to recover the
Lyapunov exponent.  The straight lines correspond to the two
asymptotic behaviors, $h(\epsilon)=h_{KS}\simeq 1.15$ and $h(\epsilon)
\sim \epsilon^{-2}$.}
\label{fig:1}
\end{figure}
%------------------------------------------------------------------------

In the following we exemplify the typical difficulties by analyzing
the map
\begin{equation}
x(t+1)= f(x(t))=x(t) + p\sin(2\pi x(t))\, .
\label{eq:mappa}
\end{equation}
As soon as $p > 0.7326\dots$, $f(x)$ is such that $f(x)>1$ and
$f(x)<0$ for some $x\in ]0,1[$. This implies that the trajectory can
travel across different unitary cells giving rise to large scale
diffusion, i.e. asymptotically:
\begin{equation}
\label{eq:coeff}
\langle [x(t)-x(0)]^2\rangle \simeq 2Dt\,,
\end{equation}
 where $D$ is the diffusion coefficient. We note that $p=O(1)$ sets
 the intrinsic scale of the displacements to be $O(1)$. Therefore, as far as 
 the $\epsilon$-entropy is concerned, for $\epsilon \ll 1$ (small scale
observations) one should be able to recognize that the system is chaotic,
i.e. $h(\epsilon)$ displays a plateau at $h_{KS}=\lambda$. For
$\epsilon \gg 1$ (large scale observations), due to the diffusive
behavior, $h(\epsilon)$ is characterized by the scaling
(\ref{eq:kolmo56}) with $\alpha=1/2$, therefore
\begin{equation}
\label{eq:3-2}
h(\epsilon) \simeq 
\left\{ 
\begin{array}{ll}
\lambda & {\rm for} \,\,\, \epsilon \ll 1 \\
D / \epsilon ^2 & {\rm for} \,\,\, \epsilon \gg 1  
\end{array}
\right.\;, 
\end{equation} 
where $\lambda$ is the Lyapunov exponent and $D$ is the diffusion
coefficient. The typical problems encountered in numerically computing
$h(\epsilon)$ can be appreciated in Fig.~\ref{fig:1}.  First notice that the
deterministic character (i.e. $h(\epsilon,\tau)\approx h_{KS}$) appears only
at $\epsilon< \epsilon_c \approx 1$. However, the finiteness of the data set
imposes a lower cut-off scale $\epsilon_d$ below which no information can be
extracted from the data (see Ref.~\cite{Olbrich97}).  As for the importance of
the choice of $\tau$ note that if $\tau$ is much larger or much shorter than
the characteristic time-scale of the system at the scale $\epsilon$, then the
correct behavior of the $\epsilon$-entropy~\cite{cenc} cannot be properly
recovered. Indeed the diffusive behavior $h(\epsilon)\sim \epsilon^{-2}$ is
roughly obtained only by considering the envelope of $h_m(\epsilon,\tau)$
evaluated at different values of $\tau$. The reason for this is that the
characteristic time of the system is determined by its diffusive behavior
$T_\epsilon \approx \epsilon^2 / D$.  On the other hand, the plateau at the
value $h_{KS}$ can be recovered only for $\tau\approx 1$, even if, in
principle, any value of $\tau$ could be used. 

We also mention that if the system is deterministic, to have a meaningful
measure of the entropy, the embedding dimension $m$ has to be larger than
information dimension of the attractor \cite{ER}.

%%%%%%%%%%%%%%%%%%%%%%%%%%%%%%%%%%%%%%%%%%%%%%%%%%%%%%%%%%%%%%%%%%%%%%
\subsubsection{Experiments on the microscopic origin of Brownian Motion}
%%%%%%%%%%%%%%%%%%%%%%%%%%%%%%%%%%%%%%%%%%%%%%%%%%%%%%%%%%%%%%%%%%%%%%
We are now ready to discuss the experiment and its analysis reported
in Ref.~\cite{gaspard}. In this experiment, a long time record (about
$1.5 \times 10^5$ data points) of the motion of a small colloidal
particle in water was sampled at regular time intervals ($\Delta
t=1/60$ s) with a remarkable high spatial resolution ($25$ nm). To our
knowledge, this is the most accurate measurement of a BM. The data
were then processed by means of standard methods of nonlinear
time-series analysis~\cite{Kantzbook} to compute the
$\epsilon$-entropy\footnote{Of course in data analysis, only
scalar time series are available and the dimensionality of the space
of state vectors is {\it a priori} unknown. However, one can use the
delay embedding technique to reconstruct the phase-space.  In this
way, the $\epsilon$-entropy can be evaluated as discussed in the
previous section. It is worth stressing that this procedure can be
applied even though the equations of motion of the system, which
generated the signal, are unknown. Moreover, this approach is meaningful
independently of the stochastic or deterministic nature of the
considered signal.}.  This computation shows a power-law dependence
$h(\epsilon)\sim \epsilon^{-2}$. Actually, similarly to what displayed
in Fig.~\ref{fig:1}, this behavior is recovered only by considering
the envelope of the $h(\epsilon,\tau)$-curves, for different $\tau$'s.
However, unlike to Fig.~\ref{fig:1}, no saturation
$h(\epsilon,\tau)\approx \mbox{const}$ is observed for small
$\epsilon$. Nevertheless, the authors {\it assume} from the outset
that the system dynamics is deterministic and, since in deterministic
systems $h(\epsilon,\tau) \leq h_{KS} \leq \sum_i^{+}\lambda_i$,
deduce from the positivity of $h(\epsilon)$ the existence of positive
Lyapunov exponents.  Their conclusion is thus that microscopic chaos
is at the origin of the macroscopic diffusive behavior.

However, as several works pointed out (see
Refs.~\cite{dettman,grass}), the huge amount of involved degrees of
freedom (Brownian particle and the fluid molecules), the impossibility
to reach a (spatial and temporal) resolution high enough, and the
limited amount of data points do not allow for such optimistic
conclusions. Avoiding a technical discussion on these three points we
simply notice that the limitation induced by the finite resolution is
particularly relevant to the experiment.  For example, if the
analysis of Fig.~\ref{fig:1} would be restricted to the region with
$\epsilon>1$ only, then discerning whether the data were originated by a
chaotic system or by a stochastic process would be impossible.
%%%%%%%%%%%%%%%%%%%%%%%% FIG.2 %%%%%%%%%%%%%%%% BEGIN
\begin{figure}[t!]
\centering
\includegraphics[draft=false, scale=0.8, clip=true]{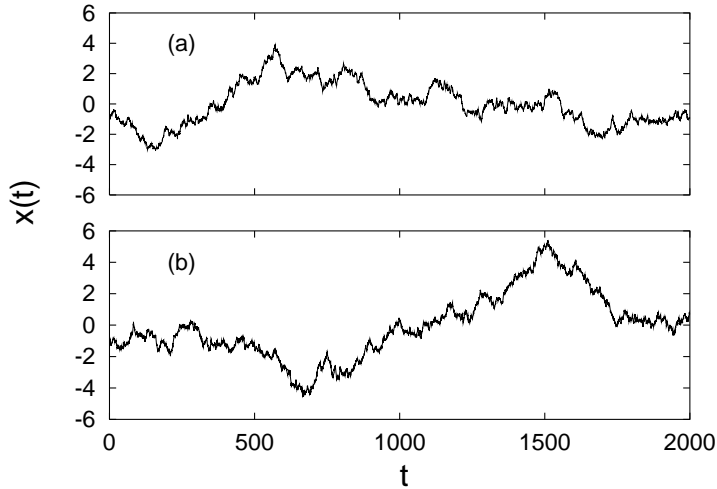}
\caption{(a) Signals obtained from Eq.~(\ref{eq:mazur}) with $M=10^4$
and random phases uniformly distributed in $[0,2\pi]$. The numerically
computed diffusion constant is $D \approx 0.007$.  (b) Time record
obtained with a continuous random walk (\ref{eq:RW}) with the same
value of the diffusion constant as in (a).  In both cases data are
sampled with $\tau=0.02$, i.e. $10^5$ data points.}
\label{fig:2}
\end{figure}
%%%%%%%%%%%%%%%%%%%%%%%%  FIG.2 %%%%%%%%%%%%%%%% END

Particularly interesting is the fact that, as shown by Dettmann et
al.~\cite{dettman,DC}, the finite amount of data severely limits our
ability to distinguish not only if the signal is deterministic, chaotic
or stochastic but also if it is deterministic regular, i.e.  of zero
entropy.  The following example serves as a clue to better understand
the way in which a deterministic non-chaotic systems may give rise (at
least on certain temporal and spatial scales) to a stochastic
behavior.

\begin{figure}[t!]
\centering
\includegraphics[draft=false, scale=0.45, clip=true]{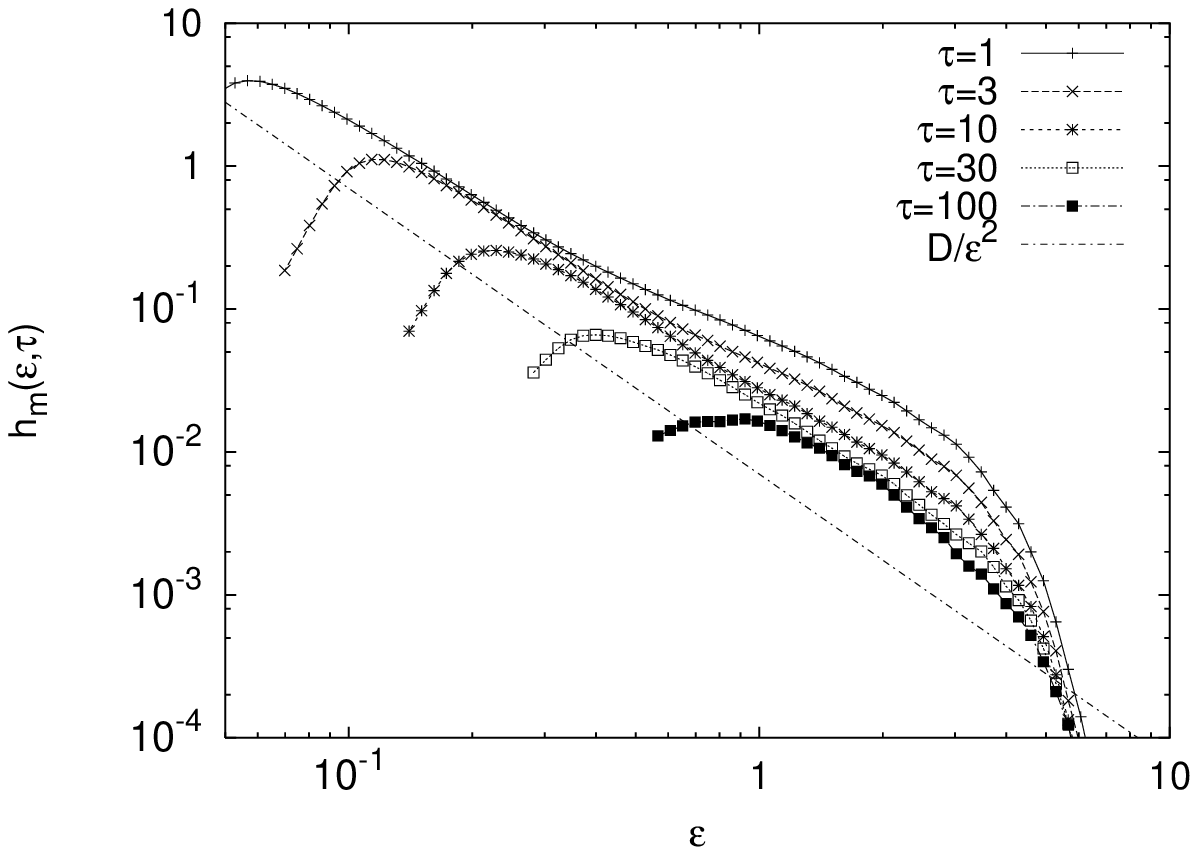}
\includegraphics[draft=false, scale=0.45, clip=true]{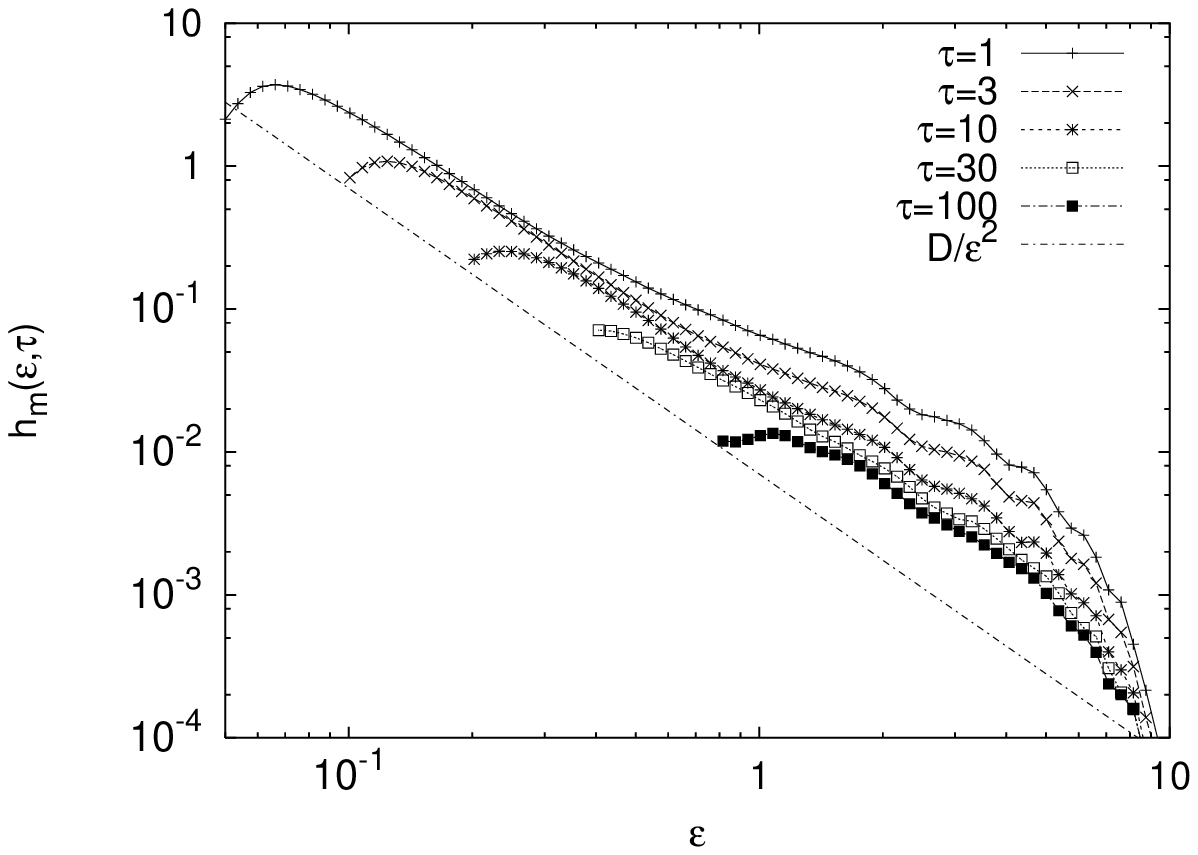}
\caption{Numerically evaluated $(\epsilon,\tau)$-entropy
using using $10^5$ points from the time series of
Fig.~\ref{fig:2}.  We show the results for embedding dimension
$m=50$.  The straight-lines show the $D/\epsilon^2$ behavior.}
\label{fig:3}
\end{figure}

Let us consider two signals, the first generated by a continuous random walk:
\begin{equation}
\dot{x}(t)=\sqrt{2D} \eta(t)\,,
\label{eq:RW}
\end{equation}
where $\eta$ is a zero mean Gaussian variable with $\langle
\eta(t)\eta(t') \rangle = \delta(t-t')$, and the second obtained as
a superpositions of Fourier modes: 
\begin{equation}
x(t)= \sum_{i=1}^M X_{0i}\sin \left(\Omega_i t+\phi_i \right) \,.
\label{eq:mazur}
\end{equation}
The coordinate $x(t)$ in Eq.~(\ref{eq:mazur}), upon properly choosing
the frequencies~\cite{mazur2,cenc} and the amplitudes
(e.g. $X_{0i}\propto \Omega_i^{-1}$), describes the motion of a heavy
impurity in a chain of $M$ linearly coupled harmonic oscillators.  We
know~\cite{mazur2} that $x(t)$ performs a genuine BM in the limit
$M\to \infty$. For $M<\infty$ the motion is quasi-periodic and
regular, nevertheless for large but finite times it is impossible to
distinguish signals obtained by (\ref{eq:RW}) and (\ref{eq:mazur})
(see Fig.~\ref{fig:2}).  This is even more striking looking at the
computed $\epsilon$-entropy of both signals (see Fig.~\ref{fig:3}).
%%%%%%%%%%%%%%%%%%%%%%%%  FIG.3 %%%%%%%%%%%%%%%% BEGIN
%%%%%%%%%%%%%%%%%%%%%%%%  FIG.3 %%%%%%%%%%%%%%%% END

The results of Fig.~\ref{fig:3} along with those by Dettman
et al.~\cite{dettman} suggest that, by assuming also the deterministic
character of the system, we are in the practical impossibility of
discerning chaotic from regular motion.

It is worth mentioning that recently some interesting works
\cite{RBastida,RBastida1} applied the entropy analysis to the motion
of a heavy impurity embedded in an FPU-chain (see
Sect.~\ref{sec:fpu}), which is a chaotic variant of the above example.
The purpose was again to infer the chaotic character of the whole
FPU-chain by observations on the the impurity motion only. It was
found that the impurity does not alter the behavior of the FPU-chain
so it can be considered as a true probe of the dynamics.  The impurity
performs a motion that, when observed at small but finite resolutions,
closely resembles a Brownian motion.  Time series
($\epsilon,\tau$)-entropy analysis both in momentum and position
allows for detecting the chaotic nature of the FPU unperturbed system,
and clearly locating the stochasticity threshold.

From the above discussion, one reaches a pessimistic view on the
possibility to detect the ``true'' nature of a signal by means of data
analysis only. However, the situation is not so bad if the question
about the character of a signal is asked only relatively to a certain
interval of scales. In this case, in fact, it is possible to give an
unambiguous classification of the signal character based solely on the
entropy analysis and free from any prior knowledge of the system/model
that generated the data.  Moreover the behavior of $h(\epsilon,\tau)$
as a function of $(\epsilon,\tau)$ provides a very useful
``dynamical'' classification of stochastic
processes~\cite{Gaspard93c,PhysicaDnostro}.  One has then a practical
tool to classify the character of a signal as deterministic or
stochastic, on a given scale, without referring to a specific model,
and is no longer obliged to answer the metaphysical question, whether
the system that produced the data was a deterministic or a
stochastic~\cite{Kubin95,cenc} one.

%%%%%%%%%%%%%%%%%%%%%%%%%%%%%%%%%%%%%%%%%%%%%%%%%%%%%%%%%%%%%%%%%
\subsection{Diffusion in deterministic non-chaotic systems}
\label{sec:nonchaotic}
%%%%%%%%%%%%%%%%%%%%%%%%%%%%%%%%%%%%%%%%%%%%%%%%%%%%%%%%%%%%%%%%%

With all the proviso on its interpretation, Gaspard's and
coworkers'~\cite{gaspard} experiment had a very positive role not only
in stimulating the discussion about the chaos/noise distinction but
also in focusing the attention on deep conceptual aspects of
diffusion.  From a theoretical point of view, the study of chaotic
models exhibiting diffusion and their non-chaotic counterpart is
indeed important to better understand the role of microscopic chaos on
macroscopic diffusion. 

In Lorentz gases, the diffusion coefficient is related, by means of
periodic orbits expansion methods~\cite{gas,dorf,rond}, to chaotic
indicators such as the Lyapunov exponents. This suggested that chaos
was or might have been the basic ingredient for diffusion. However, as
argued by Dettmann and Cohen~\cite{DC}, even an accurate numerical
analysis based on the $\epsilon$-entropy, being limited by the
finiteness of the data points, has no chance to detect differences in
the diffusive behavior between a chaotic Lorentz gas and its
non-chaotic counterpart, such as the wind-tree Ehrenfest's model.  In
the latter model, particles (wind) scatter against square obstacles
(trees) randomly distributed in the plane but with fixed orientation.
Since the reflection by the flat edges of the obstacles cannot produce
exponential separation of trajectories, the maximal Lyapunov exponent
is zero. The result of Ref.~\cite{DC} implies thus that chaos may be
not indispensable for having deterministic diffusion. The question may
be now posed on what are the necessary microscopic ingredients to
observe deterministic diffusion at large scales.

We would like to remark that, in the wind-tree Ehrenfest's
model, the external randomness amounting to the disordered  
distribution of the obstacles is crucial. Hence, one may conjecture that a
finite spatial entropy density $h_{S}$ is necessary for observing
diffusion. In this case deterministic diffusion might be a consequence
either of a non-zero ``dynamical'' entropy ($h_{KS}>0$) in chaotic
systems or of a non-zero ``static'' entropy ($h_{S} >0$) in
non-chaotic systems. This is a key-point, because someone can argue
that a deterministic infinite system with spatial randomness can be
interpreted as an effective stochastic system\footnote{This is
probably a ``matter of taste''}.

\begin{figure}
\centering
\includegraphics[draft=false, scale=0.4, clip=true]{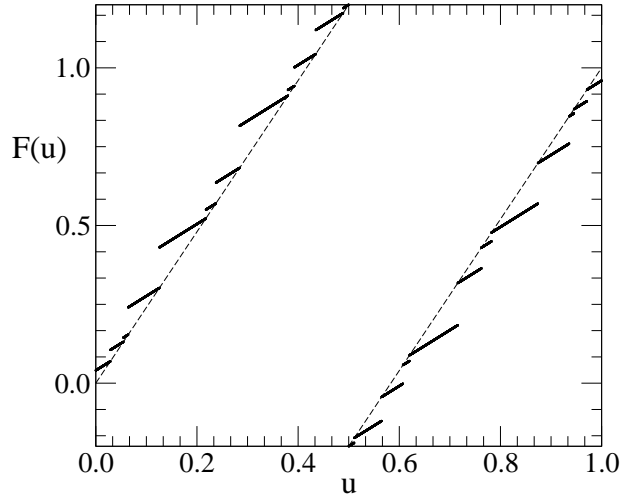}
\caption{Sketch of the random staircase map in the
unitary cell. The parameter $a$ defining the macroscopic slope is set
to $0.23$.  Half domain $[0,1/2]$ is divided into $N=12$
micro-intervals of random size.  The map on $[1/2,1]$ is obtained by
applying the antisymmetric transformation with respect to the center
of the cell $(1/2,1/2)$.}
\label{fig:4}
\end{figure}

With the aim of clarifying this point, we consider here a spatially disordered
non-chaotic model~\cite{cecco}, which is the one-dimensional analog of a
two-dimensional non-chaotic Lorentz system with polygonal obstacles.  Let us
start with the map defined by Eqs.~(\ref{eq:chaos}) and (\ref{eq:chaos2}), and
introduce some modifications to make it non-chaotic.  One can proceed as
exemplified in Fig.~\ref{fig:4}, that is by replacing the
function~(\ref{eq:chaos2}) on each unit cell by its step-wise approximation
generated as follows.
%----------------------------- FIG.4 -----------------------------------
%-----------------------------------------------------------------------
The first-half of $C_{\ell}$ is partitioned in $N$ micro-intervals $[\ell
+ \xi_{n-1},\ell + \xi_{n}[$ , $n=1,\dots, N$, with $\xi_0=0 <
\xi_1<\xi_2 < \dots <\xi_{N-1} < \xi_N=1/2$.  In each interval
the map is defined by its linear approximation
\begin{eqnarray}
F_{\Delta}(u)=
\begin{array}{ll}
u - \xi_{n} + F(\xi_n)  \qquad \quad \mbox{if~}
u \in [\xi_{n-1},\xi_{n}[ \;\; ,
\end{array}
\label{model_1}
\end{eqnarray}
where $F(\xi_n) $ is (\ref{eq:chaos2}) evaluated at $\xi_n$.  The map in the
second half of the unit cell is then determined by the anti-symmetry condition
with respect to the middle of the cell.  The quenched random variables
$\{\xi_k\}_{k=1}^{N-1}$ are uniformly distributed in the interval $[0,1/2]$,
i.e. the micro-intervals have a {\em random} extension. Further they are
chosen independently in each cell $C_{\ell}$ (so one should properly write
$\xi_{n}^{(\ell)}$). All cells are partitioned into the same number $N$ of
randomly chosen micro-intervals (of mean size $\Delta = 1/N$).  This
modification of the continuous chaotic system is conceptually equivalent to
replacing circular by polygonal obstacles in the Lorentz system~\cite{DC}.

  Since $F_{\Delta}$ has slope $1$ almost everywhere, the map is no longer
chaotic, violating the condition i) (see Sect.~\ref{sec:diffusion}).  For
$\Delta\to 0$ (i.e. $N\to \infty$) the continuous chaotic map (\ref{eq:chaos})
is recovered. However, this limit is singular and as soon as the number of
intervals is finite, even if extremely large, chaos is absent.  It has been
found~\cite{cecco} that this model still exhibits diffusion in the presence of
both quenched disorder and a quasi-periodic external perturbation
\begin{equation}
x(t+1) = [x(t)] + F_{\Delta}(x(t)-[x(t)]) + \gamma \cos(\alpha t)\,.
\label{eq:nochaos}
\end{equation}
The strength of the external forcing is controlled by $\gamma$
and $\alpha$ defines its frequency, while $\Delta$ indicates a specific
quenched disorder realization.  The sign of $\gamma$ is irrelevant; 
without lack of generality we study the case $\gamma > 0$.
%---------------------------- Fig. 5 -----------------------------------
\begin{figure}[t!]
\centering
\includegraphics[draft=false, scale=0.4, clip=true]{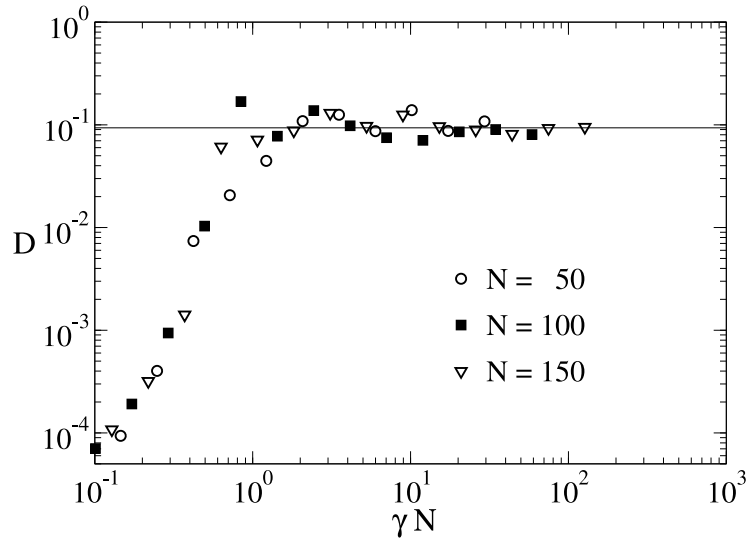}
\caption{Log-Log plot of the dependence of the diffusion coefficient
$D$ on the external forcing strength $\gamma$.  Different data
relative to a number of cell micro-intervals $N=50$, $100$ and $150$
are plotted vs the natural scaling variable $\gamma N$ to obtain
a collapse of the curves.  Horizontal line represents the result for
chaotic system~(\ref{eq:chaos},\ref{eq:chaos2}).}
\label{fig:5}
\end{figure}
%------------------------------------------------------------------------

The diffusion coefficient $D$ is then numerically computed from the
linear asymptotic behavior of the mean quadratic displacement, see
Eq. (\ref{eq:coeff}). The results, summarized in Fig.~\ref{fig:5}, show
that $D$ is significantly different from zero only for values
$\gamma>\gamma_c$.  For $\gamma> \gamma_c$, $D$ exhibits a saturation
close to the value of the chaotic system (horizontal line) defined by
Eqs.~(\ref{eq:chaos}) and (\ref{eq:chaos2}).  The existence of a
threshold $\gamma_c$ is not surprising.  Due to the staircase nature
of the system, the perturbation has to exceed the typical
discontinuity of $F_{\Delta}$ to activate the ``macroscopic''
instability which is the first step toward the diffusion. Data
collapsing, obtained by plotting $D$ versus $\gamma N$, in
Fig.~\ref{fig:5} confirms this argument.  These findings are robust
and do not depend on the details of forcing.  Therefore, we have an
example of a non-chaotic model in the Lyapunov sense by construction,
which performs diffusion.

Now the question concerns the possibility that the diffusive behavior arises 
from the presence of a quenched randomness with non zero spatial entropy 
per unit length.  To clarify this point, similarly to Ref.~\cite{DC}, 
the model can be modified in such a way that the spatial entropy per unit 
cell is forced to be zero, and see if the diffusion still persists.  
 
Zero spatial entropy per unit length may be obtained by repeating the
same disorder configuration every $M$ cells ({\em i.e.}
$\xi^{(\ell)}_n = \xi^{(\ell+M)}_n$).  Looking at the diffusion of an
ensemble of walkers it was observed that diffusion is still present
with $D$ very close to the expected value (as in Fig.~\ref{fig:5}). A
careful analysis (see Ref.~\cite{cecco} for details) showed that the
system displays genuine diffusion for a very long times even with a
vanishing (spatial) entropy density, at least for sufficiently large
$M$.

These results along with those by Dettmann and Cohen~\cite{DC} allow
us to draw some conclusions on the fundamental ingredients for observing
deterministic diffusion (both in chaotic and non-chaotic systems).
\begin{itemize}

\item An instability mechanism is necessary to ensure particle
dispersion at small scales (here small means inside the cells). In
chaotic systems this is realized by the sensitivity to the initial
condition.  In non-chaotic systems this may be induced by a finite
size instability mechanisms.  Also with zero maximal Lyapunov exponent
one can have a fast increase of the distance between two trajectories
initially close~\cite{gpt}. In the wind-tree Ehrenfest model this
stems from the edges of the obstacles, in the ``stepwise'' system of
Fig.~\ref{fig:4} from the jumps.

\item  A mechanisms able to suppress periodic orbits and therefore
to allow for a diffusion at large scale.
\end{itemize}
It is clear that the first requirement is not very strong while the
second is more subtle.  In systems with ``strong chaos'', all periodic
orbits are unstable and, so, it is automatically fulfilled.  In
non-chaotic systems, such as the non-chaotic billiards studied by
Dettmann and Cohen and the map (\ref{eq:nochaos}), the stable periodic
orbits seem to be suppressed or, at least, strongly depressed, by the
quenched randomness (also in the limit of zero spatial entropy).
However, unlike the two-dimensional non-chaotic billiards, in the 1-d
system~(\ref{eq:chaos2},\ref{model_1},\ref{eq:nochaos}), the periodic
orbits may survive to the presence of disorder, so we need the aid of
a quasi-periodic perturbation to obtain their destruction and the
consequent diffusion.

%%%%%%%%%%%%%%%%%%%%%%%%%%%%%%%%%%%%%%%%%%%%%%%%%%%%%%%%%%%%%%%%%%%%%%%%%%
\section{The heritage of the Fermi-Pasta-Ulam problem for the 
statistical mechanics}
%%%%%%%%%%%%%%%%%%%%%%%%%%%%%%%%%%%%%%%%%%%%%%%%%%%%%%%%%%%%%%%%%%%%%%%%%%

The ergodic theory begun with Boltzmann's effort to justify the
determination of average values in kinetic theory. 
Ergodic hypothesis states that time averages of observables of 
an isolated system at the equilibrium can be computed as phase averages 
over the constant-energy hyper-surface.  
This statement can be
regarded as the first attempt to establish a link between statistical
mechanics and the dynamics of the underlying system.  
One can say that
proving the validity of ergodic hypothesis provides a ``dynamical
justification'' of statistical ensembles.

The ergodic problem, at an abstract level, had been attacked by
Birkhoff and von Neumann who proved their fundamental theorems on the
existence of time averages and established a necessary and sufficient
condition for the ergodicity. In spite of their mathematical
importance, on a practical ground such theorems do not help very much
to really solve the ergodic problem in statistical physics.

There exists a point of view according to which the effectiveness of a
statistical mechanics approach resides mainly on the presence of many
degrees of freedom rather than on the underlying (chaotic or regular)
dynamics.  Khinchin in his celebrated book {\it Mathematical
Foundation of the Statistical Mechanics} \cite{khin} presents some
important results on the ergodic problem that need no metrical
transitivity.  The main point of his approach relies on the
concept of relevant physical observables in systems with a huge
number of degrees of freedom. Since physical observables are
non-generic functions (in mathematical sense), the equivalence between
time and ensemble averages should be proved only for a restricted class
of relevant observables. Moreover for physical purposes, it is
``fair'' to accept the failure of ergodicity for few (in the sense of
sets of small measure) initial conditions.

In plain words, Khinchin's formulation, coinciding with Boltzmann's
point of view (see, e.g., Ch.1 of Ref.\cite{Galla}), asserts that
statistical mechanics works, independently of ergodicity, because the
(most meaningful) physical observables are practically constant, a part
in regions of very small measure, on the constant energy  surface.  Within this
approach, dynamics have a marginal role, and the existence of ``good
statistical properties'' is granted by the large number of degrees of
freedom. However, the validity of Khinchin's statement restricts to a
special class of observables not covering all the physically
interesting possibilities. Therefore for each case, a detailed study
of the specific dynamics is generally needed to assess the statistical
properties of a given system.

The issue of  ergodicity is naturally entangled with the problem of
the existence of non-trivial conserved quantities (first integrals) in
Hamiltonian systems.  Consider a system governed by the Hamiltonian
\begin{equation}
\label{hpert}
H ( {\bm I}, \, {\bm \phi})  = H_0 ({\bm I}) + \epsilon 
H_1 ( {\bm I}, \, {\bm \phi})\, ,
\label{eq:Hpert} 
\end{equation} 
where ${\bm I}=(I_1,..., I_M)$ are the action  variables and ${\bm \phi}=(
\phi_1,..., \phi_M)$ are the phase variables.  If $\epsilon=0$ the system
is integrable, there are $M$ independent first integrals (the actions
$I_i$) and the motion evolves on $M$-dimensional tori. Two questions
arise naturally.  Do the trajectories of the system (\ref{eq:Hpert})
remain ``close'' to those of the integrable one?  Do some conserved
quantity, besides energy, survive in the presence of a generic (small)
perturbation $\epsilon H_1 ( {\bm I}, \, {\bm \phi})$?  Of course
whenever other first integrals exist the system can not be ergodic.

In a seminal work, H. Poincar\'e \cite{poincare} showed that,
generally, a system like (\ref{eq:Hpert}) with $\epsilon \neq 0$ does
not possess analytic first integrals other than energy. This result
sounds rather positive for the statistical mechanics approach.  In
1923 Fermi \cite{fermi1}, generalizing Poincar\'e's result, proved
that, for generic perturbations $H_1$ and $M>2$, there can not exist,
on the $2M-1$ dimensional constant-energy surface, even a single
smooth\footnote{For instance, analytic or differentiable enough}
surface of dimension $2M-2$ that is analytical in the variables $({\bm
  I}, \bm \phi)$ and $\epsilon$.  From this result, Fermi argued that
generic (non-integrable) Hamiltonian systems are ergodic.

At least in the physicists' community, this conclusion was generally
accepted and, even in the absence of a rigorous demonstration, there
was a vast consensus that the non-existence theorems of regular
first integrals implied ergodicity.

%%%%%%%%%%%%%%%%%%%%%%%%%%%%%%%%%%%%%%%%%%%%%%%%%%%%%%%%%%%%%%%%%%%%%%%%%%%
\subsection{FPU:  relaxation to equilibrium and ergodicity violation} 
\label{sec:fpu}
%%%%%%%%%%%%%%%%%%%%%%%%%%%%%%%%%%%%%%%%%%%%%%%%%%%%%%%%%%%%%%%%%%%%%%%%%%%
Thirty two years later Fermi itself, together with Pasta and Ulam, with one
of the first numerical experiments, in the celebrated paper {\it Studies of
non-linear problems} \cite{fpu}
(often referred with the acronym FPU) showed that the
ergodic problem was still far from being solved.
The FPU model studies the time evolution of a chain of $N$ particles,
interacting by means of non-linear springs:
\begin{equation}
H=\sum_{n=0}^{N} \left[{p^2_n\over 2m} + {K\over 2} (q_{n+1}-q_n)^2
+ {\epsilon\over \alpha} (q_{n+1}-q_n)^\alpha\right]\,,
\label{eq:fpuchain}
\end{equation}
with boundary conditions $q_0=q_{N+1}=p_0=p_{N+1}=0$, $\alpha=3$ or 4
and $K>0$.  The Hamiltonian is of the form (\ref{hpert}) with a
harmonic (integrable) part and a non-integrable (anharmonic) term
$O(\epsilon)$.  For $\epsilon=0$ one has a collection of $N$
non-interacting harmonic modes of energies $E_k$'s, which remain
constant. What happens if an initial condition is chosen in such a way
that all the energy is concentrated in a few normal modes, for
instance $E_1 (0) \neq 0$ and $E_k (0) = 0 $ for $k=2,\dots, N$?
Before the FPU work, the general expectation would have been that the
first normal mode would have progressively transferred energy to the
others and that, after some relaxation time, every $E_k(t)$ would
fluctuate around the common value.  Therefore, it came as a surprise
the fact that no tendency toward equipartition was observed, even for
long times.  In other words, a violation of ergodicity and mixing was
found.
%%%%%%%%%%%%%%%%%%%%%%%%  FIG.6 %%%%%%%%%%%%%%%% BEGIN
\begin{figure}[t!]
\centering
\includegraphics[draft=false, scale=0.5, clip=true]{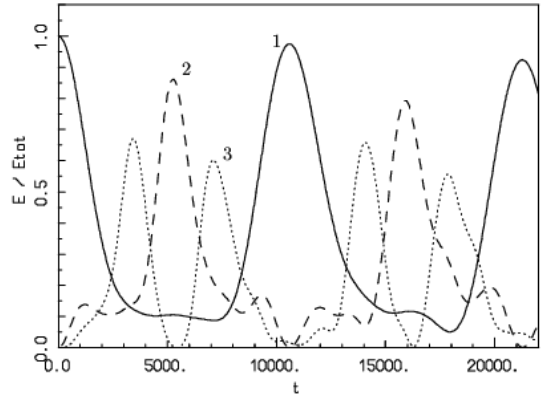}
\caption{$E_1(t)/E_{tot}$, $E_2(t)/E_{tot}$, $E_3(t)/E_{tot}$ 
for the FPU system, with $N=32$, $\alpha=3$, $\epsilon= 0.1$ and 
energy density ${\cal E} = E_{tot}/N = 0.07$. The figure is 
a courtesy of G.~Benettin~\cite{benettin})
}
\label{fig:6}
\end{figure}
%%%%%%%%%%%%%%%%%%%%%%%%  FIG.6 %%%%%%%%%%%%%%%% END
Fig.~\ref{fig:6} shows the time evolution of the fraction of energy
contained in three modes ($k=1,2,3$), in a system with $N=32$. 
%%%%%%%%%%%%%%%%%%%%%%%%% FIG. 7 %%%%%%%%%%%%%%%%%%%%%%
\begin{figure}[t!]
\centering
\includegraphics[draft=false, scale=0.5, clip=true]{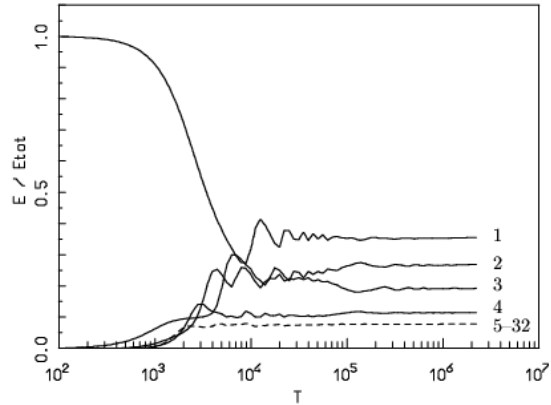}
\caption{Time averaged fraction of energy, in modes $k=1,2,3,4$
(bold lines, from top to below) and 
$ \sum _{k=5}^{32} E_{(av)k} (T)/E_{tot}$ (dashed line). 
The parameters of the system are the same as in Fig.~\ref{fig:6}.
Courtesy of G.~Benettin~\cite{benettin}.
}
\label{fig:7}
\end{figure}
%%%%%%%%%%%%%%%%%%%%%%%%%%%%%%%%%%%%%%%%%%%%%%%%
At the beginning all the energy is contained in mode $1$.  Instead of
a distribution of the energy among all the available modes, with a
loss of memory of the initial state, the system exhibits a
close to periodic behavior. The absence of equipartition can be
well appreciated looking at Fig.~\ref{fig:7}, where the quantities
\begin{equation}
\label{Emed}
E_{(av)k} (T) = \frac{1}{T} \int_{0}^{T} 
E_k(t) {\rm d} t \, , \qquad {\rm with} \quad  k=1, \dots, N \; ,
\end{equation}
i.e., the energies in the modes, averaged along the observation time
$T$, are displayed. As one can see, almost all of the energy remains
confined in the first four modes.

The existence of non ergodic behavior in non-integrable Hamiltonian
systems is actually a consequence of the so-called KAM
theorem~\cite{kolmogorov54,kamarnold,kammoser}, whose first
formulation, due to A.~N. Kolmogorov, dates back to the year before
the FPU paper.  This was surely unbeknown to Fermi and his colleagues.
The FPU result can be seen (a posteriori) as a numerical
``verification'' of the KAM theorem and, above all, of its physical
relevance, i.e. the tori survival for physically significant values of
the nonlinearity.  After Kolmogorov and FPU, it is now well
established that ergodicity is a non generic property of mechanical
systems.

Concerning the FPU problem, in terms of the KAM theorem, the following
scenario, at least for large but finite times, can be outlined
\cite{izrailev,bocchieri,benettinras}.  For $N$ particles and for a
given energy density ${\cal E} = E/N$ there is a threshold $\epsilon
_c$ for the strength of the perturbation such that
\begin{enumerate}
\item[a)] if $\epsilon < \epsilon _c $ the  KAM tori are dominant
and the system is essentially regular;
\item[b)] if $\epsilon > \epsilon _c $ the KAM  tori
are negligible and the system is essentially chaotic. 
\end{enumerate}
However, the long time evolution of very large chains with small
$\epsilon$ is hindered by the presence of metastable states. Probing
such an asymptotics by numerical simulations is extremely hard, for a
discussion on the subject see the contributions by Benettin {\it et
  al.} and Lichtenberg {\it et al.} in this volume.
%%%%%%%%%%%%%%%%%%%%%%%% FIG. 8 %%%%%%%%%%%%%%%%%%%%%%%%
\begin{figure}[t!]
\centering
\includegraphics[draft=false, scale=0.5, clip=true]{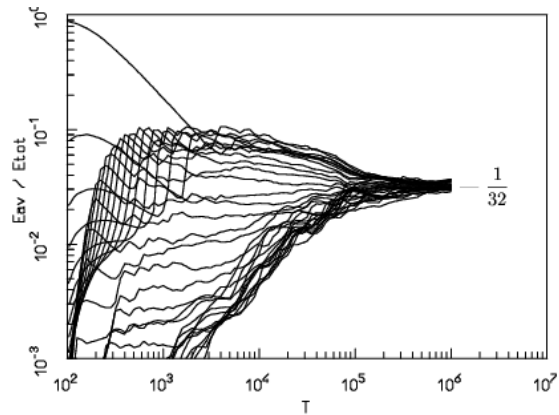}
\caption{Time averaged fraction of energy, in all the modes
$k=1,\dots,32$. The parameters of the system are $N=32$, $\alpha=3$,
$\epsilon= 0.1$ and energy density ${\cal E} = E_{tot}/N = 1.2$.
(Courtesy of G.~Benettin~\cite{benettin}).}
\label{fig:8}
\end{figure}
%%%%%%%%%%%%%%%%%%%%%%%%%%%%%%%%%%%%%%%%%%%%%%%%%%
In most of the physical situations where the strength of the perturbation
(i.e., the Hamiltonian) is fixed, the control parameter is ${\cal E}$.
There exists a critical energy density, separating regular from chaotic
behaviors. This is evident by comparing Fig.~\ref{fig:8} with
Fig.~\ref{fig:7}. In Fig.~\ref{fig:8} the same quantities of
Fig.~\ref{fig:7} are plotted, but now they refer to a system where the
energy density is much greater than before: ${\cal E} = 1.2$; the
system has entered the chaotic region and equipartition is
established.

However, also when most KAM tori are destroyed, and the system turns
out to be chaotic, the validity of ordinary statistical mechanics
is not automatically granted. Indeed the
relaxation time for reaching equipartition may become very large (see
Ref.s \cite{benettenen,kantz,casetti,deluca,ruffo,alabiso,livi} 
for a detailed discussion about this point).  

The problem of slow relaxation is rather common in high dimensional
Hamiltonian systems, where~\cite{FMV91, HO94} though the phase-space
volume occupied by KAM tori decreases exponentially with the number of
degrees of freedom (which sounds like a good news for statistical
mechanics) nonetheless very long time-scales are involved. This means
that it takes an extremely long time for the individual trajectories
to forget their initial conditions and to invade a non negligible part
of the phase space.  Indeed, even for very large systems, Arnol'd
diffusion is very weak and different trajectories, although with a
high value of the Lyapunov exponent, maintain some of their own
features for a very long time.

We conclude this part emphasizing that also in high dimensional
systems the actual role of chaos is not yet well understood. For
instance, in \cite{LPRV87} detailed numerical computations on the FPU
system show that both the internal energy and the specific heat,
computed with a time average, as functions of the temperature are
rather close to the prediction of the canonical ensemble. This is true
also in the low energy region (i.e. low temperature) where the system
behaves in a regular way (the KAM tori are dominant). This supports
Khinchin approach (though the observables are not in the class
of the sum functions\footnote{Khinchin defines sum functions as
  any function of the form $\sum_{n=1}^{N} f_n(q_n,p_n)$, $f_n$
  assuming order $1$ values. Such observables, in the large $N$ limit,
  are self-averaging, \textit{i.e.} they are practically constant
  on the constant-energy surface, aside a region of small measure.})
on the poor role of dynamics.  Indeed strong chaos seems to be
unnecessary for the prediction of the statistical mechanics to
hold. However, this is not the end of the story because in other
nonlinear systems (such as a chain of coupled rotators) the situation
is different: even in presence of strong chaos one can observe
disagreement between time average and ensemble average \cite{LPRV87}.

In the following we discuss the problem of heat transport that allows
us to discuss the role of chaos for the validity of transport
properties.

%%%%%%%%%%%%%%%%%%%%%%%%%%%%%%%%%%%%%%%%%%%%%%%%%%%%%%%%%%%%%%
\subsection{Heat transport in chaotic and non-chaotic systems}
%%%%%%%%%%%%%%%%%%%%%%%%%%%%%%%%%%%%%%%%%%%%%%%%%%%%%%%%%%%%%%

As stated in the introduction a part of the statistical mechanics
community accepts the picture according to which the instabilities of
microscopic particle dynamics are the basic requirement for the onset
of macroscopic transport.  In this framework, several works
\cite{gas,dorf} have shown that, in some systems, there exists a
relationship between transport coefficients (thermal or electrical
conductivity, viscosity, diffusivity etc.) and Lyapunov exponents.
Such a link is of remarkable importance because it establishes a
straightforward connection between the microscopic dynamical
properties of a system and its macroscopic behavior, which is the
  main goal of statistical mechanics.  However, as exemplified in the
  previous sections, chaotic dynamics does not seem to be a necessary
  condition to both equilibrium and out-of-equilibrium statistical
  mechanics approaches. In fact, we have seen that transport may occur
  even in the absence of deterministic chaos. These counterexamples
  pose some doubts on the generality and so on the conceptual
  relevance of the links found between chaotic indicators and
  macroscopic transport coefficients.

Heat conduction is a typical phenomenon that needs a microscopic
mechanism leading to normal diffusion that distributes particles and
their energy across the whole system. Since a chaotic motion has the
same statistical properties of a ``random walk'', when observed at
finite resolution, this mechanism can be found in the presence of
either exponential instability in deterministic dynamics or intrinsic
disorder and nonlinearities.

\begin{figure}
\begin{center}
\includegraphics[draft=false, scale = 0.5, clip=true]{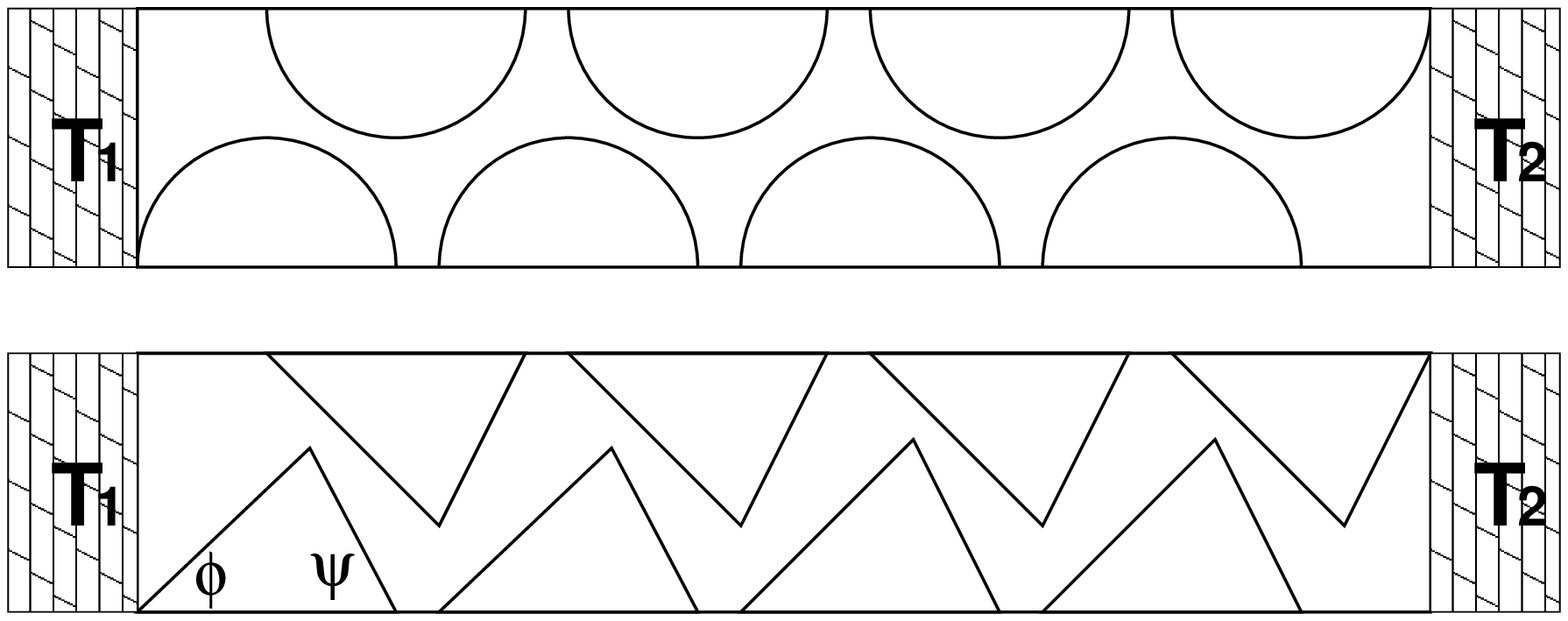}
\end{center}
\caption{Example of channel geometry used in 
Ref.~\cite{Alonso,li2} to study
heat transport in low dimensional chaotic (upper panel) and non chaotic 
billiards (lower panel).}
\label{fig:channel}
\end{figure}

In the context of the conduction problem, FPU chains have recently played an
important role in further clarifying the transport properties of low spatial
dimension systems. FPU models represent simple but non trivial candidates to
study heat transport by phonons in solids whenever their boundaries are kept
at different temperatures. This issue becomes even more interesting at low
spatial dimensions where the constraints set by the geometry may induce
anomalous transport properties characterized by the presence of divergent
transport coefficients in the thermodynamic limit \cite{LLP03}. Thermal
conductivity $\chi$, defined via the Fourier's Law
$$
J = -\chi \nabla T, 
$$ 
relates the heat (energy) flux $J$ to a temperature
gradient.  When a small temperature
difference $\delta T = T_1 - T_2$ is applied to the ends of a system of
linear size $L$, the heat current across the system is expected to be
$$
J = - {\chi \delta T \over L} \,\, .
$$ For some one- and even two-dimensional systems, theoretical
arguments, confirmed by several simulations, predict a scaling
behavior $J \sim L^{\alpha -1}$ implying a size dependent conductivity
\begin{equation}
\chi(L) = L^{\alpha}\;. 
\end{equation}
As a consequence, $\chi$ diverges in the limit $L\to \infty$ with a
power law whose exponent $\alpha>0$ depends on the specific system
considered.  The presence of this divergence is referred to as {\it
anomalous heat conduction} in contrast with normal conduction which,
according to dimensional analysis of Fourier's Law, prescribes a
finite limit for $\chi$.  FPU chains are systems where the anomaly in
the heat transport is clearly observed. Its origin can be traced back
to the existence of low-energy modes which survive long enough to
propagate freely before scattering with other modes. Such modes can
carry much energy and since their motion is mainly ballistic rather
than diffusive, the overall heat transport results to be anomalous.
Models other than FPU indeed presents this peculiar conduction, as
widely shown in the literature \cite{LiWangHu,li2,GNY}. Then the issue
is the general understanding of the conditions leading to this phenomenon and
more specifically the role of microscopic dynamical instabilities.  A
well know chaotic system, such as the Lorentz Gas in a channel
\cite{Alonso} configuration, provides an example of a system with
normal heat conduction. This model consists of a series of
semicircular obstacles with radius $R$ arranged in a lattice along a
slab of size $L\times h$ ($h<<L$) see Fig.~\ref{fig:channel}.   As in a
Lorentz system, particles scatter against obstacles but do not
interact with each other.

%------------------------------------------------------------------
%------------------------------------------------------------------

Two thermostats at temperatures $T_1$ and $T_2$ respectively are placed at
each end of the slab to induce transport. They reinject into the system those
particles reaching the ends with a velocity drawn from a Gaussian velocity
distribution with variance proportional to the temperatures $T_1$ and $T_2$.
In the case of semicircular obstacles the system is chaotic and one 
observes a standard Fourier's Law \cite{Alonso}.

In references \cite{LiWangHu} and \cite{li2}, some non-chaotic variants of the 
Lorentz channel have been proposed in order to unravel the role of
exponential instabilities in the heat conduction. In those models,
called the Ehrenfest Channel, the semicircular obstacles were replaced 
with triangular ones, so that the system is trivially non-chaotic since
collisions with flat edges of the obstacles cannot separate
trajectories more than algebraically. The results show that 
when two angles ({\textit e.g.} $\phi$ and $\psi$) of the triangles are 
irrational multiple of $\pi$, the system exhibits a
normal heat conduction. On the contrary, for rational ratio, such as  
isosceles right triangles, the conduction becomes anomalous. 
The single particle heat flux across $N$
cells $J_1(N)$ scales as $J_1(N) \sim N^{\alpha}$, while the
temperature gradient behaves as $1/N$ implying that $\chi(N)$ diverges as
$N\to \infty$.  The explanation of such a divergence can be found in
the single-particle diffusivity along the channel direction which
occurs in a non standard way. Indeed, the evolution of a large set of 
particles
has a mean squared displacements from initial conditions which grows in 
time with a power law behavior
$$
\langle [x(t) - x(0)]^2 \rangle \sim  t^b
$$ with an exponent $1<b<2$.  
This super-diffusion is the unique responsible for a
divergent thermal conductivity independently of 
Lyapunov instabilities, since the model has a zero Lyapunov exponents.

When an Ehrenfest Channel with anomalous thermal conductivity is 
disordered, for instance, by randomly
modulating the height of triangular obstacles or their positions along
the channel, the conduction follows Fourier's law, becoming 
normal~\cite{LiWangHu}.
This scenario is rather similar to that one discussed in Sect.~\ref{sec:nonchaotic} 
for diffusion on non chaotic maps.

The works in Refs.~\cite{LLP03,LiWang,Zhao} suggest that the anomalous
conduction is associated with the presence of a mean free path of
energy carriers that can behave abnormally in the thermodynamic limit.
For FPU the long mean free path is due to soliton-like ballistic
modes. In the channels the long free flights, between consecutive
particle collisions, become relevant.  The above considerations
suggest a very week role of chaos for heat transport, and for 
transport in general, since also systems without exponential 
instability may show transport, even anomalous.

%%%%%%%%%%%%%%%%%%%%%%%%%%%%%%%%%%%%%%%%%%%%%%%%%%%%%%%%%%%%%%%%%%%%%%%
\section{Concluding remarks}
%%%%%%%%%%%%%%%%%%%%%%%%%%%%%%%%%%%%%%%%%%%%%%%%%%%%%%%%%%%%%%%%%%%%%%%

The problem of distinguishing chaos from noise cannot receive an
absolute answer in the framework of time series analysis.  This is due
to the finiteness of the observational data set and the impossibility
to reach an arbitrary fine resolution and high embedding
dimension. However, we can classify the signal behavior, without
referring to any specific model, as stochastic or deterministic on a
certain range of scales.
 
Diffusion may be realized in both stochastic and deterministic 
systems.  In particular, as the analysis of polygonal 
billiards and non-chaotic maps (see sect.~\ref{sec:nonchaotic}) shows, 
chaos is not a prerequisite for observing diffusion and, more in general, 
nontrivial statistical behaviors.

In a similar way, we have that for the validity of heat conduction
chaos is not a necessary ingredient. Also in systems with zero maximal
Lyapunov exponent (see ref.s \cite{Alonso,li2,GNY}) the Fourier's law
(or its anomalous version) can hold.

We conclude by noticing that the poor role of exponential instability
for the validity of statistical laws does not seem to be limited to
transport problems.  For instance it is worth mentioning the
interesting results of Lepri et al.~\cite{lepri} showing that the
Gallavotti-Cohen formula~\cite{galla}, originally proposed for chaotic
systems, holds also in some non-chaotic model.

\subsubsection*{Acknowledgments}
The authors express their gratitude to 
D. Del-Castillo-Negrete, O.~Kantz and E.~Olbrich who recently 
collaborated with them on the issues discussed in 
this chapter. A special thank to G.~Benettin for having provided us 
with figures \ref{fig:6}, \ref{fig:7}, \ref{fig:8}. 

\backmatter%%%%%%%%%%%%%%%%%%%%%%%%%%%%%%%%%%%%%%%%%%%%%%%%%%%%%%%
%%%%%%%%%%%%%%%%%%%%%%%%%%%%%%%%%%%%%%%%%%%%%%%%%%%%%%%%%%%%%%%%%%

%%%%%%%%%%%%%%%%%%%%%%%%%%%%%%%%%%%%%%%%%%%%%%%%%%%%%%%%%%%%%%%%%%%%%%

\end{document}